\documentclass[preprint,5p,times,twocolumn]{elsarticle}

\usepackage{amssymb}
\usepackage{textcomp} 

\usepackage{subcaption} 
\usepackage{hyperref} 
\usepackage{listings} 
\usepackage{xcolor} 
\usepackage{amsmath}  
\usepackage[acronym]{glossaries} 
\usepackage{enumitem} 
\glsdisablehyper 
\newacronym{caf}{CAF}{Coarray Fortran}
\newacronym{gpu}{GPU}{graphics processing unit}
\newacronym{pgas}{PGAS}{partitioned global address space}
\newacronym{gpgpu}{GPGPU}{general purpose computing on \gls{gpu}}
\newacronym{cpu}{CPU}{central processing unit}
\newacronym{ofi}{OFI}{Open Fabric Interface}
\newacronym{rma}{RMA}{remote memory access}
\newacronym{uma}{UMA}{uniform memory access}
\newacronym{numa}{NUMA}{non-uniform memory access}
\newacronym{spmd}{SPMD}{single program multiple data}
\newacronym{mpi}{MPI}{Message Passing Interface}
\newacronym{fft}{FFT}{Fast Fourier Transform}
\usepackage{booktabs}
\usepackage{multirow}
\usepackage{tabularx}

\journal{Journal of Parallel and Distributed Computing}

\begin{document}

\lstset{
  basicstyle=\ttfamily,
  columns=fullflexible,
  breaklines=true,
  postbreak=\mbox{$\hookrightarrow$\space},
}

\begin{frontmatter}



\title{Accelerating Fortran Codes: A Method for Integrating Coarray Fortran with CUDA Fortran and OpenMP} 


\author[inst1,inst2]{James McKevitt}
\ead{james.mckevitt@univie.ac.at}
\author[inst1]{Eduard I. Vorobyov}
\author[inst3]{Igor Kulikov}

\affiliation[inst1]{organization={University of Vienna, Department of Astrophysics},
            addressline={Türkenschanzstrasse 17}, 
            city={Vienna},
            postcode={A-1180}, 
            country={Austria}}

\affiliation[inst2]{organization={University College London, Mullard Space Science Laboratory},
            addressline={Holmbury St Mary}, 
            city={Dorking},
            postcode={RH5 6NT}, 
            state={Surrey},
            country={United Kingdom}}

\affiliation[inst3]{organization={Institute of Computational Mathematics and Mathematical Geophysics SB RAS},
            addressline={Lavrentieva ave. 6}, 
            city={Novosibirsk},
            postcode={630090}, 
            country={Russia}}

\begin{abstract}
Fortran's prominence in scientific computing requires strategies to ensure both that legacy codes are efficient on high-performance computing systems, and that the language remains attractive for the development of new high-performance codes. Coarray Fortran (CAF), part of the Fortran 2008 standard introduced for parallel programming, facilitates distributed memory parallelism with a syntax familiar to Fortran programmers, simplifying the transition from single-processor to multi-processor coding. This research focuses on innovating and refining a parallel programming methodology that fuses the strengths of Intel Coarray Fortran, Nvidia CUDA Fortran, and OpenMP for distributed memory parallelism, high-speed GPU acceleration and shared memory parallelism respectively. We consider the management of pageable and pinned memory, CPU-GPU affinity in NUMA multiprocessors, and robust compiler interfacing with speed optimisation. We demonstrate our method through its application to a parallelised Poisson solver and compare the methodology, implementation, and scaling performance to that of the Message Passing Interface (MPI), finding CAF offers similar speeds with easier implementation. For new codes, this approach offers a faster route to optimised parallel computing. For legacy codes, it eases the transition to parallel computing, allowing their transformation into scalable, high-performance computing applications without the need for extensive re-design or additional syntax.
\end{abstract}



\begin{keyword}
Coarray Fortran (CAF) \sep CUDA Fortran \sep OpenMP \sep MPI
\PACS 0000 \sep 1111
\MSC 0000 \sep 1111
\end{keyword}

\end{frontmatter}



%
%
%
%

\section{Introduction}

Across the many fields which make use of scientific computing, the enduring importance of Fortran-written codes is undeniable. Most notably, Intel's compiler remains a popular choice for these Fortran codes, primarily due to its robust performance and reliable support. However, with the exponential growth in computational demands, there is an imperative need to enhance the speed and efficiency of these codes. Shared memory parallelism techniques, like OpenMP, though useful and easy to implement, often fall short in meeting these demands. Hence, turning to distributed memory parallelism and graphics processing units (GPUs) becomes essential, given their capacity to exploit modern and computationally efficient hardware \cite{Brodtkorb2013GraphicsComputing,Owens2008GPUComputing}.

To optimise the use of GPUs in general computing tasks (general purpose computing on GPUs; GPGPU; \cite{Santander-Jimenez2019ComparativeParsimony,Khairy2019AHeterogeneity,Owens2007AHardware}), Nvidia's CUDA, a parallel computing platform and programming model, is often employed \cite{Garland2011NVIDIAGPU,Che2008ACUDA}. Fortran users can leverage CUDA Fortran, an adapted language also provided by Nvidia, which offers all the speed advantages of CUDA, but with the familiar Fortran syntax \cite{GregRuetsch2012AnFortran,Ruetsch2014CUDAEngineers,NVIDIA2023CUDAGuide}. The true potential of CUDA Fortran is unlocked when applied to tasks that involve heavy parallelisation like Fast Fourier Transform (FFT) operations \citep[e.g.,][]{VanDeWiele2014FastImplementation}, often a common and performance-critical component in astrophysics simulations and image or data processing \citep[e.g.,][]{Binney1987GalacticDynamics,Vorobyov2023ComputingMethod}.

For distributed memory parallelism the Message Passing Interface (MPI) is commonly used \cite{Gropp2011MPIInterface}. However, its implementation can be resource-intensive and often requires a full re-write of the original serialised code. We turn, however, to Coarray Fortran, as a simpler yet powerful alternative \cite{Sharma2017MPIMeshes,Garain2015ComparingApplications}. Coarray Fortran, introduced in the Fortran 2008 standard, is designed specifically for parallel programming, both with shared and distributed memory \cite{Numrich1998Co-arrayProgramming}. It not only offers a simple syntax but also ensures efficient performance, especially in the Intel implementation, easing the transition from single-to-multiple-node programming.

The fusion of these two paradigms offered by separate providers, while non-trivial, offers a powerful combination to accelerate Fortran codes on the most modern hardware using intuitive Fortran syntax. This paper provides a comprehensive guide on how to perform such an acceleration of Fortran codes using CUDA Fortran and Coarray Fortran. Regardless of the specific problem at hand, our methodology can be adapted and implemented by scientists in their computational work. We demonstrate the advantages and drawbacks of various available configurations, using a well-established and highly parallelisable potential field solver as a case study \cite{Vorobyov2023ComputingMethod}. We also explore various techniques and strategies, such as the usage of pointers to optimise communication speed and implement direct memory access. Additionally, we delve into the definition of variables required by both the central processing unit (CPU) and GPU memory, highlighting the treatment of variables describing a potential field in our case study as an example.

Our proposed approach has a broad focus, providing a roadmap that can be adapted to any Fortran code. Through this detailed guide, we aim to enable researchers to streamline their computational workflows, augment their codes' speed, and thereby accelerate their scientific work.

\subsection{Coarray Fortran}
Coarray Fortran (CAF), introduced in the Fortran 2008 standard, has gained recognition for its ability to simplify parallel programming. CAF offers an intuitive method for data partitioning and communication, enabling scientists to focus on problem-solving rather than the intricacies of parallel computing.

Coarray Fortran integrates parallel processing constructs directly into the Fortran language, eliminating the need for external libraries like MPI. It adopts the Partitioned Global Address Space (PGAS) model, which views memory as a global entity but partitions it among different processors \cite{Almasi2011PGASLanguages}. This allows straightforward programming of distributed data structures, facilitating simpler, more readable codes.


In CAF, the global computational domain is decomposed into a series of images, which represent self-contained Fortran execution environments. Each image holds its local copy of data and can directly access data on other images via coarray variables. These variables are declared in a similar fashion to traditional Fortran variables but with the addition of codimensions. The codimension has a size equal to the number of images. The variables defined without codimensions are local to each image.


Coarray Fortran introduces synchronisation primitives, such as \texttt{critical}, \texttt{lock}, \texttt{unlock}, and \texttt{sync} statements, to ensure proper sequencing of events across different images. This provides programmers with the tools to manage and prevent race conditions and to coordinate communication between images.

\subsection{CUDA Fortran}
CUDA Fortran is a programming model that extends Fortran to allow direct programming of Nvidia GPUs. It is essentially an amalgamation of CUDA and Fortran, providing a pathway to leverage the computational power of Nvidia GPUs while staying within the Fortran environment.

The CUDA programming model works on the basis that the host (CPU) and the device (GPU) have separate memory spaces. Consequently, data must be explicitly transferred between these two entities. This introduces new types of routines to manage device memory and execute kernels on the device.

In CUDA Fortran, routines executed on the device are known as kernels. Kernels can be written in a syntax very similar to standard Fortran but with the addition of qualifiers to specify the grid and block dimensions. This is how CUDA Fortran harnesses the power of GPUs - by organising threads into a hierarchy of grids and blocks around which the hardware is constructed. 

An essential aspect of CUDA Fortran programming is understanding how to manage memory effectively, which involves strategically allocating and deallocating memory on the device and copying data between the host and the device. Special attention needs to be given to optimising memory usage to ensure that the full computational capability of the GPU is used.

CUDA Fortran can be integrated seamlessly with existing Fortran codes, offering a less labour-intensive path to GPU programming than rewriting codes in another language \cite{Ruetsch2014CUDAEngineers}.

\subsection{Heterogeneous Computing with CPUs and GPUs}

When developing a code capable of running on a heterogeneous system --- that being a system containing more than one type of processor, in this case a CPU-GPU system --- it is important to understand the differences in the architecture of parallel execution between the hardware to best distribute tasks and optimize code performance.

Single Instruction, Multiple Data (SIMD) is a parallel computing model where one instruction is executed simultaneously across multiple data elements within the same processor \citep{Flynn1972SomeEffectiveness}. This process is known as vectorisation, and it allows a single processor to perform the same operation on multiple data points at once.

In GPU architectures, SIMD is combined with multi-threading and implemented in a way where warps (groups of threads) receive the same instruction. This means that while multiple threads perform identical operations in parallel with each other, each thread also performs vectorised (SIMD) operations on its assigned data. This broader GPU parallelism is known as Single Instruction, Multiple Threads (SIMT) \citep{McCool2012StructuredComputation}.

While each GPU thread is capable of performing hundreds or thousands of simultaneous identical calculations through vectorisation, individual CPU cores also support SIMD but on a smaller scale, typically performing tens of operations simultaneously each. However, unlike in GPUs, each core within a CPU multiprocessor can receive its own set of instructions. This means that while individual cores perform SIMD operations, the overall CPU executes multiple instructions across its multiple cores, following the Multiple Instruction, Multiple Data (MIMD) model \citep{Flynn1972SomeEffectiveness}.


This means that besides the difference in scale of simultaneous operations between CPUs and GPUs, there are also architectural differences in how CPU MIMD and GPU SIMT handle parallel tasks. In MIMD architectures like those in CPUs, each core operates independently, allowing more flexibility when encountering conditional branching such as an \texttt{if} statement. This independence helps minimise performance losses during thread divergence because each core can process different instructions simultaneously without waiting for others.

Conversely, in GPU architectures using SIMT, all threads in a warp execute the same instruction. This synchronization can lead to performance bottlenecks during thread divergence —-- for instance, when an \texttt{if} statement causes only some threads within a warp to be active. In such cases, GPUs typically handle divergence by executing all conditional paths and then selecting the relevant outcomes for each thread, a process that can be less efficient than the CPU’s approach. This synchronisation requirement makes GPUs highly effective for large data sets where operations are uniform but potentially less efficient for tasks with varied execution paths. Thus, while GPUs excel in handling massive, uniform operations due to their SIMD capabilities within each thread, CPUs offer advantages in scenarios where operations diverge significantly.

It is, therefore, important to understand which problems are best suited to which type of processor, and design the code to distribute different tasks to the appropriate hardware.

The layout of this work is as follows. In Sect. 2 we outline our methodology and consider a number of options which are available to solve this problem, depending on the use case. We also present a detailed guide on how compiler linking can be achieved. In Sect. 3 we present the results of our tests on different hardware. In Sect. 4 we summarise our main conclusions.

\section{Methodology}
The methodology we propose hinges on the robust combination of CUDA Fortran and Coarray Fortran, leveraging their unique strengths to develop efficient high-performance computing applications. The primary challenge is the complex interfacing required to integrate the GPU-accelerated capabilities of Nvidia CUDA Fortran with the distributed memory parallelism of Intel Coarray Fortran. CUDA Fortran is chosen as it allows the user high levels of control over GPU operations - some of which are highlighted below - while remaining close to the traditional Fortran syntax. Likewise, Coarray Fortran, particularly when implemented by Intel, allows for high distributed memory performance with no augmentations to the standard Fortran syntax.
Below, we detail the steps involved in our approach.

\subsection{Selection of Compilers}
CUDA Fortran, being proprietary to Nvidia, requires the use of Nvidia's \texttt{nvfortran} compiler. However, \texttt{nvfortran} does not support Coarray Fortran. In contrast, Intel's \texttt{ifort} compiler supports Coarray Fortran - with performance levels that rival MPI and without the complexity of its syntax - but does not support CUDA Fortran. One other alternative compiler supporting Coarray Fortran is \texttt{OpenCoarrays}. According to our experience, its implementation falls short in terms of speed. However, we note that experiences with OpenCoarrays may vary depending on a variety of factors such as hardware configurations and compiler versions.

When considering most Fortran programmes in general, Intel's \texttt{ifort} compiler is a common choice and offers a high-performance, robust and portable compiler. Consequently, here we demonstrate a hybrid \texttt{ifort} for Coarray Fortran and \texttt{nvfortran} for CUDA Fortran solution.

\subsection{Memory Space Configuration}

When creating a single code which requires the use of two compilers, a few key considerations are required. For the following text, \lq{}Intel code\rq{} refers to code compiled with \texttt{ifort} and \lq{}Nvidia code\rq{} refers to code compiled with the \texttt{nvfortran}. The various execution streams mentioned below refer to an executed command being made within either the Intel code or the Nvidia code.

\subsubsection{Pageable and Pinned Memory}

Before continuing, a short background on pageable and pinned memory is required.
Pageable memory is the default memory type in most systems. It is so-called because the operating system can \lq{}page\rq{} it out to the disk, freeing up physical memory for other uses. This paging process involves writing the contents of the memory to a slower form of physical memory, which can then be read back into high-speed physical memory when needed.

The main advantage of pageable memory is that it allows for efficient use of limited physical memory. By paging out memory that is not currently needed, the operating system can free up high-speed physical memory for other uses. This can be particularly useful in systems with limited high-speed physical memory.

However, the paging process can be slow, particularly when data is transferred between the host and a device such as a GPU. When data is transferred from pageable host memory to device memory, the CUDA driver must first allocate a temporary pinned memory buffer, copy the host memory to this buffer, and then transfer the data from the buffer to the device. This double buffering incurs overhead and can significantly slow down memory transfers, especially for large datasets.

Pinned memory, also known as page-locked memory, is a type of memory that cannot be paged out to the disk. This means that the data is constantly resident in the high-speed physical memory of the system, which can result in considerably faster memory transfers between the host and device.

The main advantage of pinned memory is its speed. Because it can be accessed directly by the device, it eliminates the need for the double-buffering process required by pageable memory on GPUs. This can result in significantly faster memory transfers, particularly for large datasets.

However, pinned memory has its drawbacks. The allocation of pinned memory is more time-consuming than pageable memory - something of concern if not all memory allocation is done at the beginning - and it consumes physical memory that cannot be used for other purposes. This can be a disadvantage in systems with limited physical memory. Additionally, excessive use of pinned memory can cause the system to run out of physical memory, leading to a significant slowdown as the system starts to swap other data to disk. Pinned memory is not part of the Fortran standard and is, therefore, not supported by the Intel compiler. This leads to additional intricacies during the combination of compilers, as we discuss below.

\subsubsection{Host memory: Coarray Fortran}

Returning to our problem, when using Intel code and Nvidia code together, the division of parameters and variables between the two is one of the key areas to which attention should be paid. The execution stream within the Nvidia code can only access and operate on variables and parameters which have been declared in the Nvidia code. Likewise, the Intel execution stream can only access and operate on variables and parameters which have been declared in the Intel code. For this reason, we consider a virtual partition within the host physical memory, which can be crossed through the use of subroutines and interfaces, which we discuss in more detail below. This clear division of the physical memory does not happen in reality, as Intel and Nvidia declared variables and parameters will be mixed together when placed in the physical memory, but it is a helpful tool to consider the possible configurations we detail below.

\begin{figure}
  \centering
  \includegraphics[width=\columnwidth]{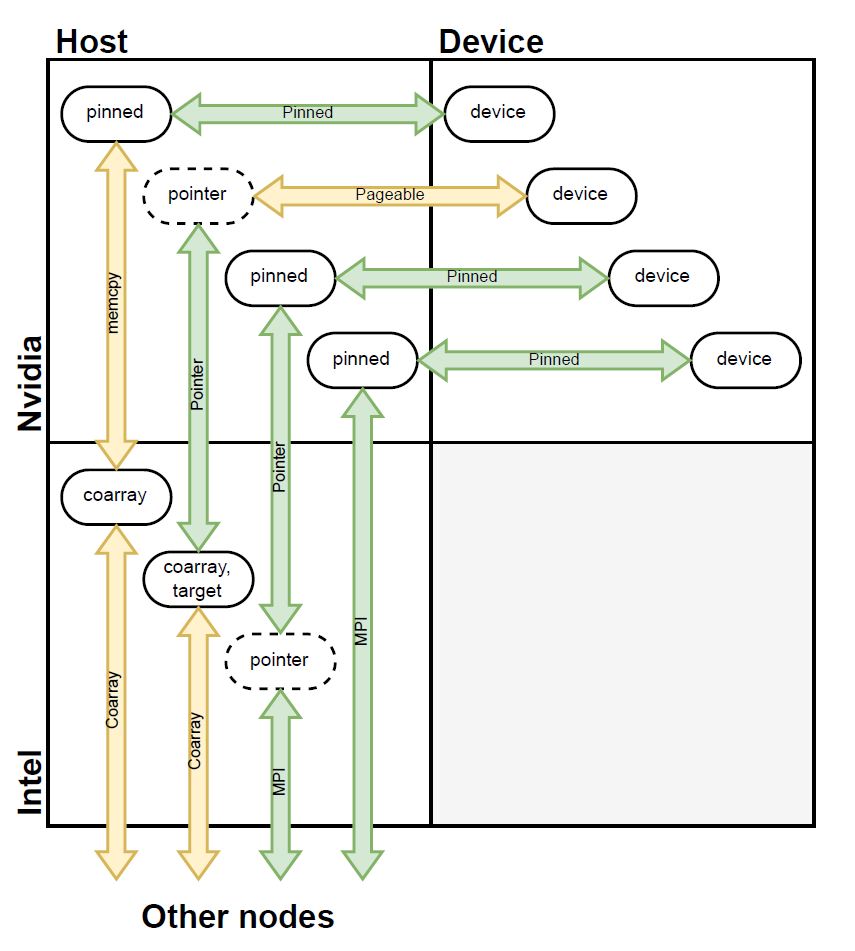}
  \caption{The relative transfer speeds of moving variables between various parts of the host memory and the device memory, where green represents a fast to negligible speed, and yellow represents a slower speed. As CUDA Fortran is used for GPU operations, only Nvidia code can be used to define device variables. MPI is added for comparison.}\label{fig:possibilities}
\end{figure}

Depending on which side of the partition in the host memory the variables are placed defines where and at what speed they can be transferred. Figure~\ref{fig:possibilities} shows a selection of these options. In the leftmost case, a variable has two identical copies in the host memory, one defined by the Intel compiler to allow CAF parallelisation (simply using the \texttt{[*]} attribute) and one by the Nvidia compiler to allow it to be pinned. Proceeding to the right, in the next case, the variable is defined by the Intel compiler for CAF communication, but a pointer to this variable is defined by the Nvidia compiler, meaning it cannot be pinned and so suffers from the slower pageable memory transfers to the GPU. In the next cases, we consider the options available with MPI. The variable is defined by the Nvidia compiler which allows it to be pinned, but a pointer to this variable is defined by the Intel compiler. This means that CAF transfers are not possible as they are not supported for pointers, meaning MPI is required for distributed memory parallelism. In the final case, the Intel code is not required at all, as the Nvidia compiler supports these MPI transfers natively. To understand any overhead associated with the pointing procedure which allows the combination of CAF and CUDA, we implemented the two MPI solutions to our potential solver and found no appreciable difference in performance. Any small speedup or slowdown between the two options --- one using pointers to combine both compilers and one only using the Nvidia compiler --- is likely due to the different MPI versions and implementations used in the compilers, rather than by the pointing procedure.


The partition-crossing use of such a variable is much slower when copying the array than when using a pointer, which has practically no speed overhead at all. In the case that the variable is copied across the partition, the version which is defined in the Nvidia code can be defined with attributes allowed by the Nvidia compiler, one of these being that it is pinned. This would not be possible in the Intel version of the variable because, as previously discussed, pinned memory is not part of the Fortran standard and so is not supported by the Intel compiler. It is important to note that, as mentioned above, the Nvidia compiler natively supports MPI, and can therefore run Fortran codes parallelised across shared memory, distributed memory and GPUs. While Coarray Fortran offers a simpler syntax, it requires a slightly more complex compilation process and setup when using GPU parallelisation. We, therefore, lay out this process as clearly as possible in this article, to make the simplified speed-up of Coarray Fortran as accessible as possible.


The pointer solution, while allowing a faster cross-partition solution, does not allow for this pinned attribute, as the Nvidia code simply points to the array defined by the Intel compiler, which resides in pageable memory. This means a pageable transfer rather than a pinned transfer takes place when moving the values onto the device, which is slower.

It is important to think about the problem which is being addressed as to which solution is more optimal. On our hardware and testing with a 128$^3$ array, pinned memory transfers took around 1 ms, as opposed to 3 ms taken by pageable memory transfers. However, during cross-partition operations, using a pointer resulted in approximately $\sim$0 ms of delay, as opposed to 3 ms when using a copy operation. Given the choice between 1) 1 ms GPU transfers but 3 ms cross-partition transfers, or 2) 3 ms GPU transfers but $\sim$0 ms cross-partition transfers, it is important to consider the specific use case.

In our case, and therefore also in most cases, it is more optimal to use a pointer in the Nvidia code. This does not mean pinned memory transfers to the device cannot be used at all. They simply cannot be used for any coarrayed variables, which reside in the Intel code partition. During implementation, it is also clearly easier to implement the pointer configuration, which we describe now.

\begin{figure*}
    \centering
    \includegraphics[width=\textwidth]{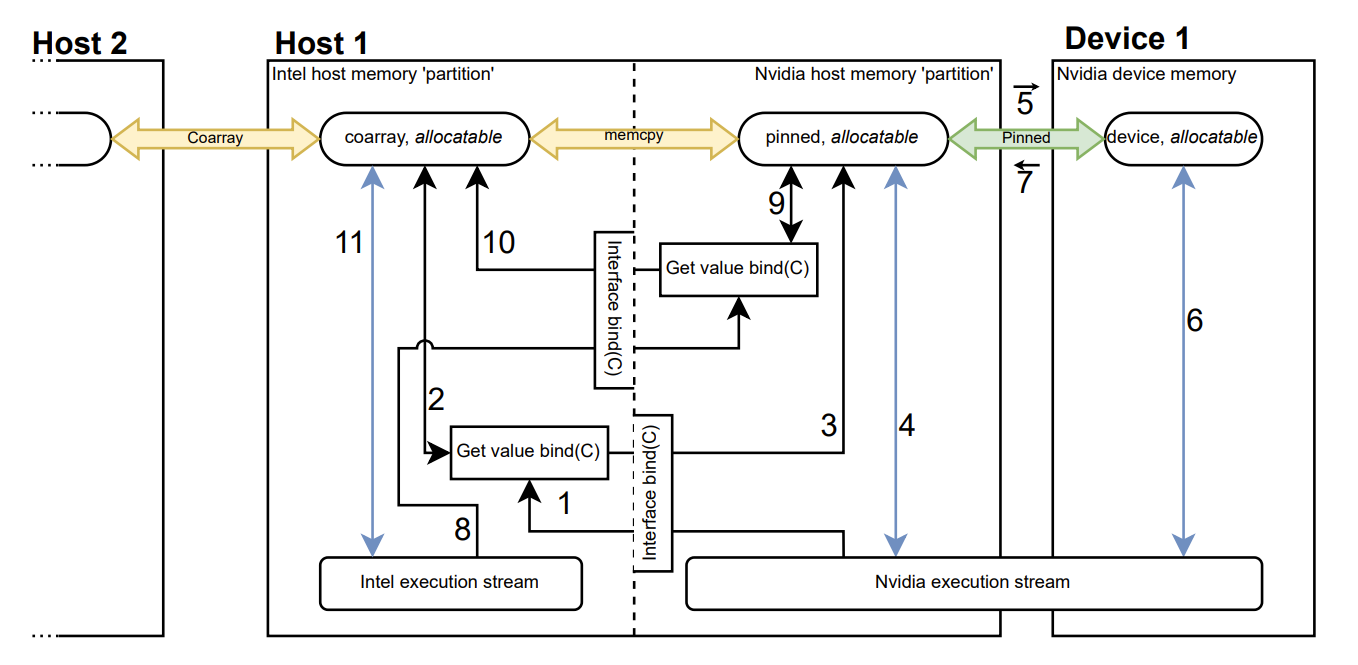}
    \caption{Configuration of physical memory and the procedure required for retrieval across the partition allowing pinned memory for fast device transfers but slow partition crossing by copying memory. In this case, a variable has been changed by the Intel execution stream before this procedure starts, and so the Nvidia execution stream is required to retrieve an updated copy of this array before operating on it, transferring it to the device and performing further GPU-accelerated operations. The Intel execution stream then takes over, retrieving the value(s) of the updated variable, performing its operations, and then allowing for transfer to other host memory spaces in the distributed memory. This host-host transfer must occur every time a device operation is performed with a coarrayed variable.}
    \label{fig:config_slow}
\end{figure*}

First, we show the implementation of two copies of the same array stored in the physical memory of the host, either side of our partition.  We note that this is inefficient and makes poor use of the memory available given that parameters and variables requiring transfer are stored twice, although this is not typically a problem for modern high-performance computing systems. Furthermore, every time a variable is updated by the execution stream of one side of the code, a transfer is required across the host physical memory causing some considerable slowdown --- particularly in the case of large arrays which are frequently updated. Technically, a transfer could only be made if it is known that the data will be changed on the other side of the partition before being used again, but this introduces a large and difficult-to-detect source for coding errors, especially in complex codes, and so is not advisable. This transfer could also be performed asynchronously to mask the transfer time, but the inefficient use of available physical memory is intrinsic to this solution. An illustration of how this setup works can be seen in Figure~\ref{fig:config_slow}, the stages of which are as follows:

\begin{enumerate}[align=left]
  \item[1.] The Nvidia execution stream calls a subroutine, which has been defined and compiled within the Intel code, to get the value of the array (in this case a coarray), which is to be operated on. To do this, an interface is defined within the Nvidia code with the \texttt{bind(C)} attribute, meaning that while the source code is Fortran, the compiled code is C, ensuring a robust connection between the two compilers. The destination subroutine is also written with the \texttt{bind(C)} attribute for the same reason. An example subroutine showing how to use C-binding is provided in the appendix.
  \item[2--3.] The subroutine returns the value of the array and sets the value of the counterpart array in the Nvidia code to the same value.
  \item[4.] The array is now updated on the Nvidia partition and can be operated on by the Nvidia execution stream.
  \item[5.] A pinned memory transfer can now take place, moving the array to the device memory.
  \item[6.] The array is operated on by the device.
  \item[7.] A pinned memory transfer allows the array to be moved back to the Nvidia partition of the host memory.
  \item[8--11.] An inverse of the first four steps takes place, allowing the updated array to be used by the Intel execution stream, and accessible for coarray transfer to the rest of the distributed memory programme.
\end{enumerate}

What was found to be more efficient in our use case, and in general something which is true for the majority of cases in terms of execution speed, memory usage and coding simplicity, was to declare a parameter or variable once in either the Intel or Nvidia code, and then create a pointer to this parameters or variable in the partition of the code. This means that whenever, for example, the Intel execution stream requires a large array from the Nvidia code, no slowdown is caused during the transfer of the array to the other partition and no unnecessary overhead in the physical memory is present.

\begin{figure*}
    \centering
    \includegraphics[width=\textwidth]{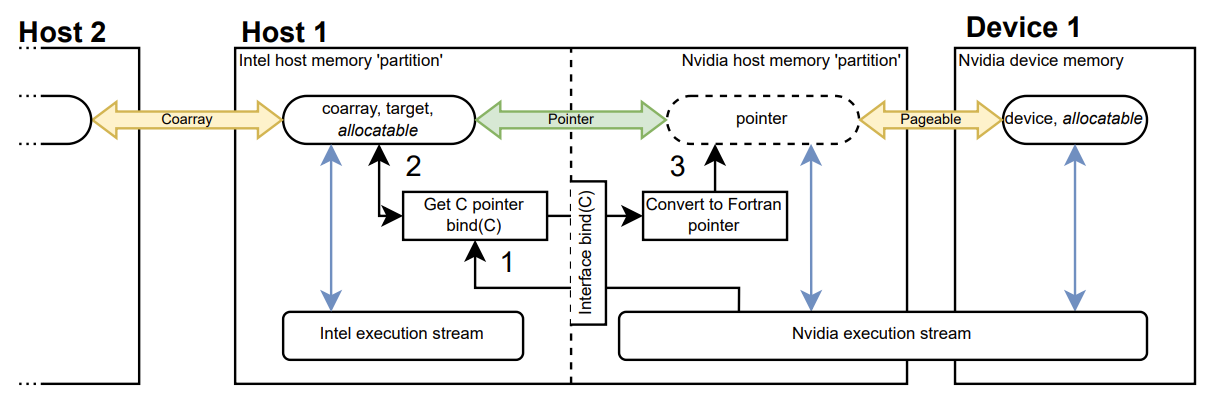}
    \caption{Configuration of physical memory and the procedure required for retrieval across the partition using pointers allowing for no transfer overhead in host-host transfers, but requiring the use of pageable memory for device transfers and incurring an overhead. In this case, the linking is performed only once during the first stages of the running of the code, meaning both host execution streams are always accessing the same version of host variables.}
    \label{fig:config_fast}
\end{figure*}

This pointing can be accomplished using an intermediate series of subroutines, calls, and pointers, all of which are written in Fortran but declared with the c-binding to ensure a robust and common connection between the two compilers, as mentioned above. In the prior duplication case, these subroutines and calls are required before the use of any parameters or variables which have been changed by the alternate execution stream since their last use, to ensure an up-to-date version is used. However, with this new pointer case, the subroutines and calls are only required once at the beginning of the running to setup and initialise the pointing, making it harder to mistakenly forget to update an array before using it in the source code. This configuration can be seen in Figure~\ref{fig:config_fast}. The steps of this configuration, to set up a coarrayed variable capable of distributed memory transfer and a counterpart pointer which allows transfer to the device for GPU accelerated operations, are as follows:

\begin{enumerate}[align=left]
  \item[1.] The Nvidia execution stream again calls a subroutine, defined and compiled within the Intel code, to get the address in the memory of the array (in this case a coarray), which is to be operated on. This must be declared with the target attribute in the Intel code, so it can be pointed to. As before, an interface is defined within the Nvidia code with the \texttt{bind(C)} attribute, to allow this connection. The destination subroutine is also written with the \texttt{bind(C)} attribute, and the pointer is made a \texttt{C} pointer for the same reason.
  \item[2.] The subroutine returns the address of the array as a C-pointer.
  \item[3.] The C-pointer is converted to a Fortran pointer for use by the Nvidia code. The original coarray, as defined by the Intel code and residing in the Intel partition of the host memory, can now be operated on and transferred to the device for GPU-accelerated operations.  We note that in this case the data are transferred from the pageable host memory to the device, thus incurring a certain overhead, as described above. 
\end{enumerate}

This handling of sensitive cross-compiler operations with C-bindings was found to be essential, as relying on Fortran-Fortran interfacing between the compilers led to non-descript segmentation errors. The use of C-binding ensures a robust solution to this issue, with no overhead in the execution speed and no deviation from the Fortran syntax in the source code.

\subsubsection{Host memory: MPI}

\begin{figure*}
  \centering
  \includegraphics[width=\textwidth]{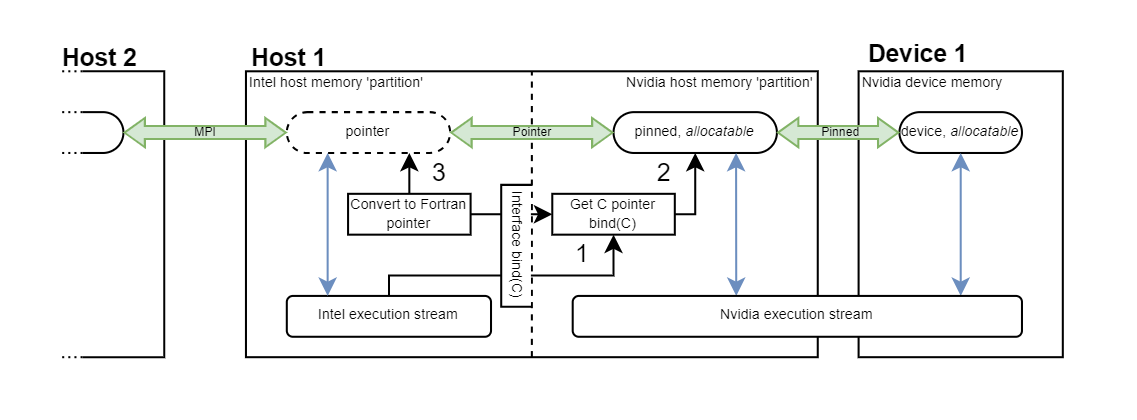}
  \caption{Configuration of physical memory and the procedure required for retrieval across the partition using pointers allowing for no transfer overhead in host-host transfers, and allowing for the use of pinned memory for device transfers, incurring the minimal overhead. As in the previous case, the linking is performed only once during the first stages of the running of the code, meaning both host execution streams are always accessing the same version of host variables.}
  \label{fig:config_mpi}
\end{figure*}

This technique is almost identically applicable when using MPI. However, it should be noted that in that case, an additional benefit is present in that arrays which are communicated using the MPI protocol do not require a coarray attribute in their definition. This means that, opposite to the case in Figure~\ref{fig:config_fast}, the array can be defined in the Nvidia code, and the pointer in the Intel code.  In this case, the full speed-up of data transfer for the pinned memory can be utilized. An illustration of this can be seen in Figure~\ref{fig:config_mpi}.

\subsection{CPU Affinity with GPUs}

In any high-performance computing application using either PGAS, or more widely the single program multiple data (SPMD) technique, the ability to control the allocation of programme processes to specified processing units---commonly referred to as CPU affinity or process pinning---is pivotal for performance optimisation. This is especially relevant in systems with a Non-Uniform Memory Access (NUMA) architecture (common in HPC), and/or in systems with multiple CPUs and GPUs.

\begin{figure}
  \centering
  \includegraphics[width=.5\linewidth]{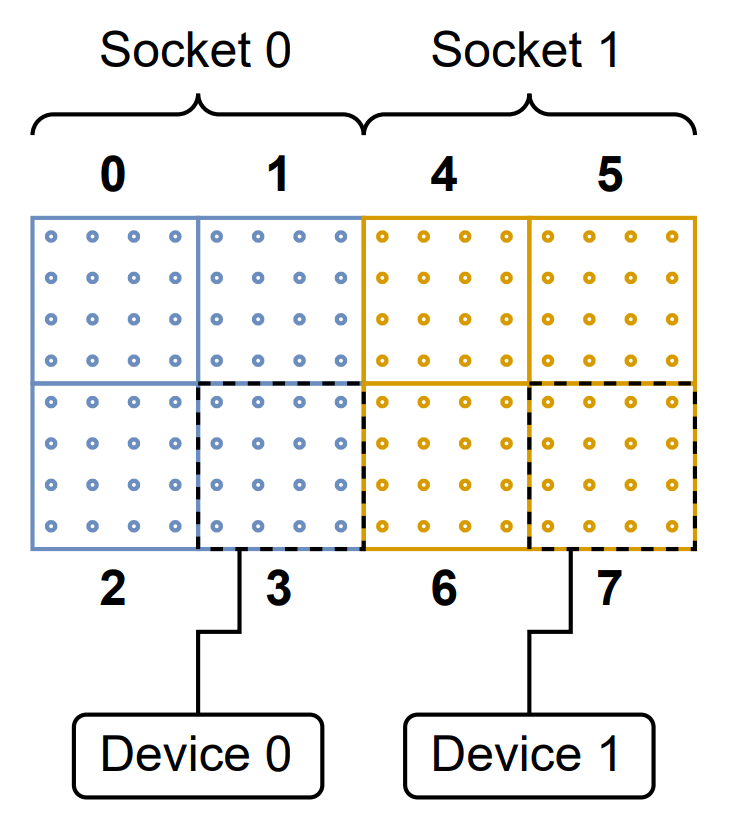}
  \caption{Architectural schematic detailing the interconnect between GPUs and NUMA nodes within a dual-socket configuration.}
  \label{fig:numa_gpu}
\end{figure}

\begin{figure}
  \centering
  \includegraphics[width=.9\linewidth]{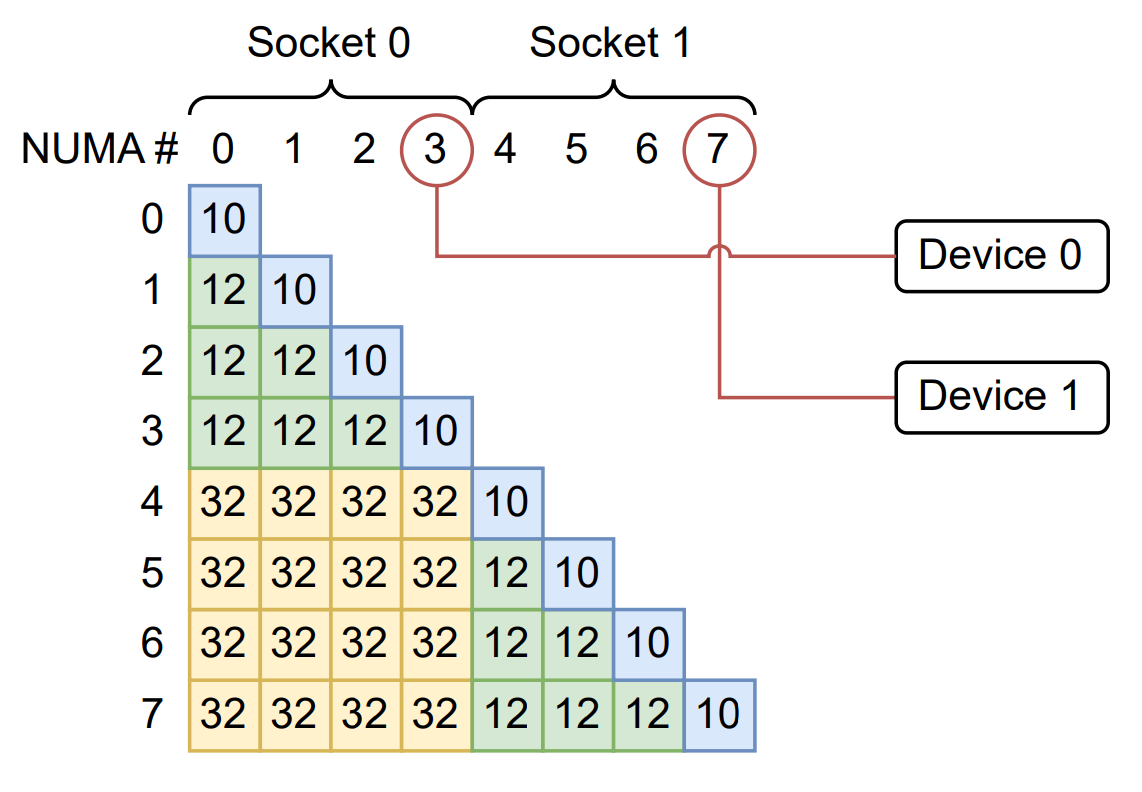}
  \caption{Schematic representation of NUMA node distances within a multi-socket, multi-GPU environment. NUMA nodes closely connected to the GPUs are emphasised.}
  \label{fig:distances}
\end{figure}

In a NUMA architecture, the term \lq{}NUMA node\rq{} refers to a subsystem that groups sets of processors within one multiprocessor together and connects them with a memory pool, forming a closely-knit computational unit. Unlike a Uniform Memory Access (UMA) configuration where each processor within the multiprocessor has equal latency for memory access across the system, NUMA nodes have differing latencies depending on whether the accessed memory resides within the same NUMA node or in a different one. Taking the example of one of our systems, described in detail later, this can be seen in Figures~\ref{fig:numa_gpu} and~\ref{fig:distances}. When using a hybrid programming model, such as a PGAS language like Coarray Fortran in combination with OpenMP in our case, it becomes specifically important to remember these.

Therefore, a well-configured system strives to keep data and tasks localised within the same NUMA node whenever possible, mitigating the impact of varying memory access latencies across NUMA nodes. This concept is critical for understanding the intricacies of CPU affinity and process pinning in multi-socket, multi-GPU systems.

The Linux command \texttt{numactl~--hardware} can be used to determine relative distances between NUMA nodes, the memory available to each group, and the ID of each processor within each NUMA node. If more exact latency numbers are required between NUMA nodes, tools like the Intel Memory Latency Checker can be used. If hyperthreading is enabled on a system, then additional processor IDs will be shown in addition to those referring to physical processors. In our case, our first NUMA node, NUMA node 0, encompasses the processors 0--15 and 128--143, with the former referring to the physical processors and the latter referring to their hyperthreaded counterparts. We do not concern ourselves with hyperthreading in this paper, as we observed no speed improvements when using it.

A second important aspect of the NUMA architecture when considering GPUs is the connection of GPUs to the processors, found using the Linux command \texttt{nvidia-smi~topo~--matrix}. In our architecture, a GPU is connected to one NUMA node, with one device connected per socket. As seen in Figure~\ref{fig:distances}, each NUMA node within a socket has a slight speed penalty associated with communication with NUMA nodes in the same socket (20\% increase) and a high overhead associated with communicating with the other socket (220\% increase). When a coarray image on one socket, therefore, attempts to communicate with the GPU of the other, there is a big overhead in the communication. Where our GPUs are connected is shown in the figure.

For practical application within Intel Coarray Fortran, the Intel MPI Library offers environment variables to facilitate CPU affinity settings:

\begin{verbatim}
export I_MPI_PIN=on
export I_MPI_PIN_DOMAIN=[HEX_CONFIG]
\end{verbatim}
\noindent{}where \texttt{HEX\_CONFIG} is a hexadecimal (hex) code corresponding to the IDs of the compute cores within the multiprocessors which are to be connected to the GPUs. This is described in more detail below.

To effectively set CPU affinity through Intel MPI's environment variables, one must provide the correct hexadecimal (hex) codes corresponding to the CPU or core IDs. These hex codes serve as unique identifiers for the CPUs and are critical for pinning computational tasks accurately.
In the case of our code, each coarray image is designed to require one GPU. It is, therefore, important to ensure that each coarray image is running on the socket containing the NUMA node that is connected to the correct GPU, namely, to the GPU the coarray image is using. Other situations where multiple coarray images use the same GPU are also possible, in which case these images should be pinned within the same socket the GPU is connected to. However, if more images are added to a socket, they each naturally must reside on fewer cores, meaning that any CPU parallelisation is limited. Determining the balance between CPU and GPU parallelisation is something that will vary between use cases.
As aforementioned, in our example, we are interested in CPU affinity because we want to ensure that each coarray image runs on the socket containing the NUMA node that is connected to the GPU it is using.

Once the desired destination CPU IDs have been identified, they can be converted to hexadecimal format in the following way: Given two coarray images are running on one node (in this case, one image on each socket), the CPU code of the first image should be pinned to NUMA node 3 and CPU code of the second image to NUMA node 7, with the corresponding GPU code pinned to GPUs 0 and 1 respectively. The GPU number used by the coarray image can simply be set using the Nvidia CUDA Fortran line

\begin{lstlisting}
istat = cudaSetDevice(mod(irank-1,gpusPerNode))
\end{lstlisting}

\noindent{}where \texttt{irank} is the Coarray Fortran image number, starting from 1 and \texttt{gpusPerNode} is the number of GPUs attached to each computational node, this being 2 in the case shown in Figures~\ref{fig:numa_gpu} and~\ref{fig:distances}. This will effectively alternate between 0 and 1 as the image number (\texttt{irank}) increases from 1 to the last image.

Taking our test case, it is known that NUMA node 3 corresponds with cores 48--63. This first needs to be encoded in binary, considering all the 127 cores we have available:

\begin{lstlisting}
00000000000000000000000000000000 00000000000000001111111111111111 00000000000000000000000000000000 00000000000000000000000000000000
\end{lstlisting}

This is then converted to a hex code:
\texttt{FFFF0000000000000000}.
The same is done for NUMA node 7 (cores 112-127) and the appropriate hex code (\texttt{FFFF}), is generated. Once obtained, these hex codes can be placed in the \texttt{I\_MPI\_PIN\_DOMAIN} environment variable to set CPU affinity precisely as:

\begin{lstlisting}
export I_MPI_PIN=on
export I_MPI_PIN_DOMAIN=[FFFF0000000000000000,FFFF]
\end{lstlisting}

\texttt{I\_MPI\_PIN\_DOMAIN} here receives two arguments for positions within the node (\texttt{[position 1, position 2]}), given that in our case we want to pin two coarray images to each node. Each coarray image is then allocated, starting with the first image (1) and proceeding in order to the last image, to position 1 on node 1, position 2 on node 1, position 1 on node 2, position 2 on node 2, and so on.

Activating these environment variables and running the Coarray Fortran executable adheres the computation to the designated CPU affinity. The fidelity of this configuration can be corroborated via the \texttt{I\_MPI\_DEBUG} environment variable.
The performance difference between pinning coarray images on the correct, and incorrect, sockets for their associated GPUs, can be seen in our scaling tests below.

\subsection{Compiling and Linking}

Finally, we compile the CUDA Fortran code with \texttt{nvfortran} into constituent object files and then into one shared object library file. The Coarray Fortran code is compiled with \texttt{ifort} into constituent object files. To link the two together, we use \texttt{ifort}, ensuring that the linking is performed in an environment where both CUDA Fortran and Coarray Fortran libraries are included in the library path. We outline a simplified procedure here:
\begin{enumerate}
    \item Compile the CUDA Fortran device code into a position-independent machine code object file:
    \newline\texttt{nvfortran -cuda -c -fpic device.cuf}
    \item Create a shared object library file using the object file:
    \newline\texttt{nvfortran -cuda -shared -o libdevice.so device.o}
    \item Compile the Fortran host code into a machine code object file:
    \newline\texttt{ifort -c host.f90}
    \item Create a CAF configuration file:
    \newline\texttt{echo \textquotesingle{}-n 2 ./a.out\textquotesingle{} > config.caf}
    \item Link the host machine code object file and the device machine code shared object library file, also using CAF distributed memory flags:
    \newline\texttt{ifort -coarray=distributed -coarray-config-file=config.caf -L. host.o -ldevice -o a.out}
\end{enumerate}

Before execution of the programme, the relative path to the source directory should also be appended to the dynamic shared libraries path list environment variable \texttt{LD\_LIBRARY\_PATH}.

Our methodology involves a meticulous combination of CUDA Fortran's GPU acceleration and Coarray Fortran's efficient data distribution across multiple compute nodes. It requires careful attention to compiler selection, memory allocation, and interfacing between two different programming paradigms. The result is a seamless integration of these distinct models into a single high-performance computing application with high-performance GPU and distributed memory acceleration.

\subsection{Intel Coarray Fortran with AMD processors}

Intel's \texttt{ifort} does not officially support the running of code on AMD processors. Therefore, there are inevitably some complications when attempting to do this. Many high-performance computing centres use AMD processors, including the ones used by us for this study. We faced errors when running our code on AMD processors from the remote memory access (RMA) part of Coarray Fortran (CAF). Unlike pure-MPI, CAF uses a one-side communications protocol which means that instead of each image being told to send and receive data, one image is allowed to access the data from another without any action from the target image. CAF still uses MPI commands, and therefore requires an MPI version, but only uses the one-sided protocols.

We used the \texttt{ifort} compiler in \texttt{intel/2019} as it was the most recent version of the intel compiler available to us, and found that it cannot perform such RMA operations when AMD processors are being used, and therefore CAF programmes cannot run. There are three solutions we identified to this problem, the first of which is to change the version of MPI which is used by \texttt{ifort} to one released by \texttt{OpenMPI}. The second is to change the Open Fabric Interface (OFI) version. The third, and in our experience the fastest, solution is to change the internal default transport-specific implementation. This still allows RMA operations to take place but through a slightly different, AMD processor-compatible, method.

This can be done by setting the environment variable \texttt{MPIR\_CVAR\_CH4\_OFI\_ENABLE\_RMA=0}. This is an environment variable setting for the Message Passing Interface (MPI) implementation, specifically for the MPICH library, which is a widely used open-source MPI implementation. The Intel Fortran compiler supports MPI for parallel programming, and the MPICH library can be used in conjunction with the Intel Fortran compiler for this purpose. The environment variable \texttt{MPIR\_CVAR\_CH4\_OFI\_ENABLE\_RMA} is related to the CH4 device layer in MPICH, which is responsible for communication between processes. CH4 is a generic communication layer that can work with different network interfaces, and OFI (Open Fabric Interface) is one such interface. The \texttt{MPIR\_CVAR\_CH4\_OFI\_ENABLE\_RMA} variable controls the usage of RMA operations in the CH4 device layer when using the OFI network interface. RMA operations enable processes to directly access the memory of other processes without involving the target process, thus providing low-latency communication and better performance.

\begin{figure}
    \centering
    \begin{subfigure}{\columnwidth}
        \centering
        \includegraphics[width=.8\linewidth]{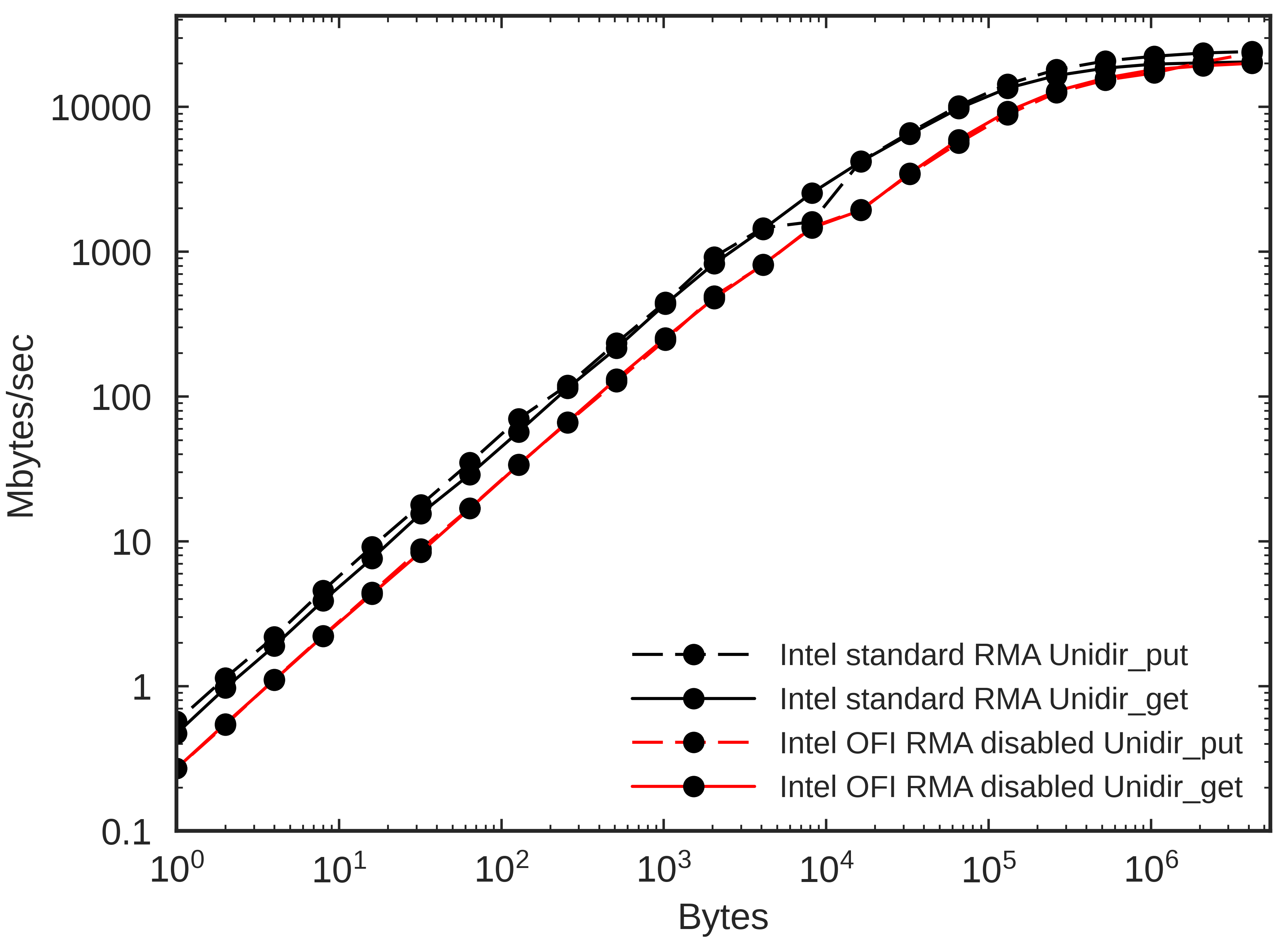}
        \caption{Entire data range from 0 to 10$^7$ bytes shown.}
    \end{subfigure}
    \hfill
    \begin{subfigure}{\columnwidth}
        \centering
        \includegraphics[width=.8\linewidth]{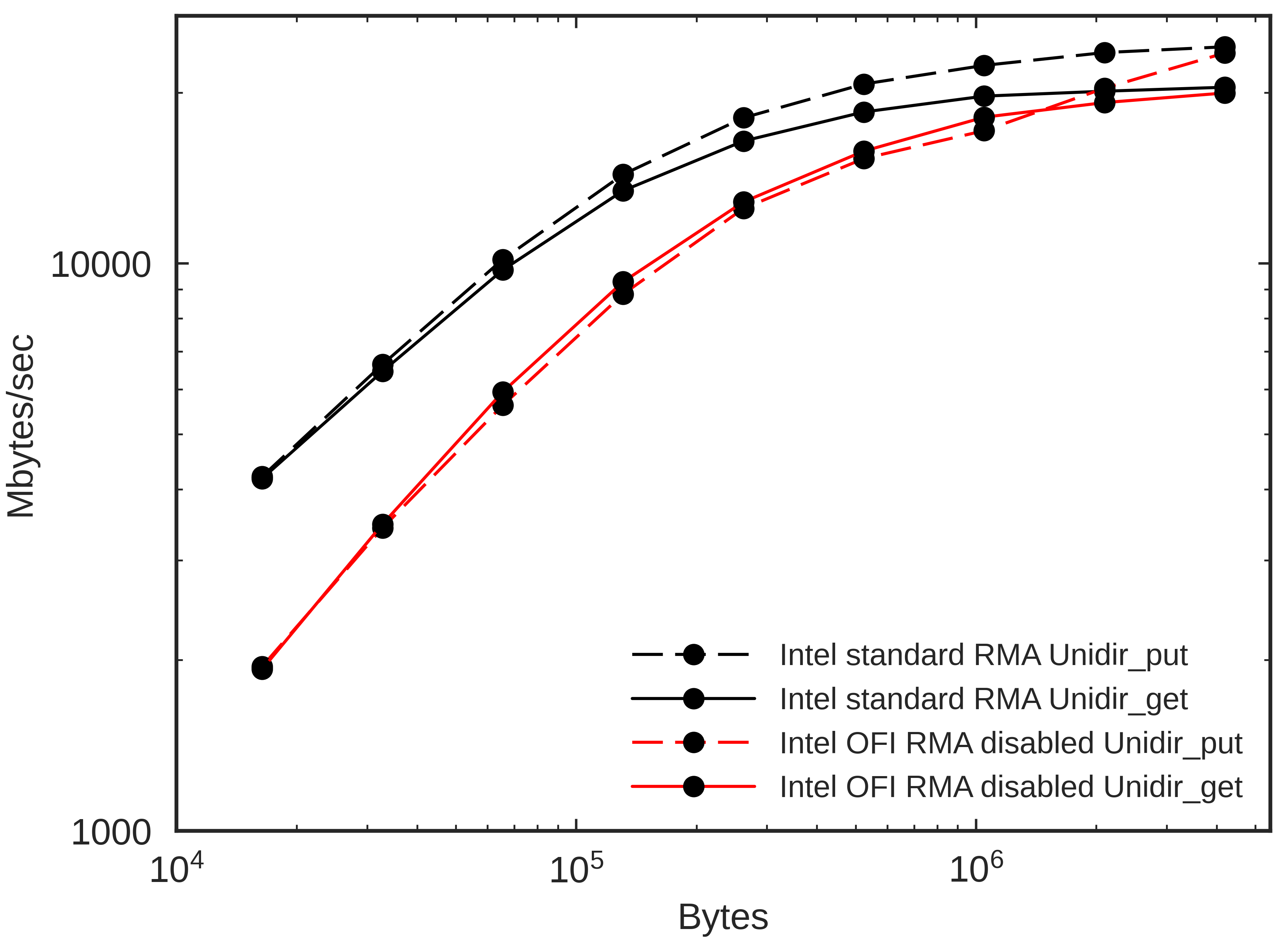}
        \caption{Data range cropped to that between 10$^4$ to 10$^6$ bytes shown.}
    \end{subfigure}
    \caption{Results of one-sided memory transfers between two images on separate nodes both with and without the adapted RMA procedure, showing a slight speed overhead.}
    \label{fig:ofi_rma}
\end{figure}

To understand the overhead associated with changing the internal default transport-specific implementation, we ran speed tests using the Intel Benchmarking tool on some of the MPI operations which are used by CAF both with and without OFI RMA disabled. We used the \texttt{unidir\_put} and \texttt{unidir\_get} commands for this comparison, the results of which can be seen in Figure~\ref{fig:ofi_rma}. As is seen in this Figure, there is little effect above a few~$\times 10^6$~bytes, saturating to no speed overhead incurred by changing the RMA implementation as we describe. Therefore, our comparisons of speeds between MPI and our slightly augmented CAF can be seen as representative of a direct comparison between MPI and CAF.

\section{Performance Analysis and Results}

\begin{table*}[h]
  \centering
  \begin{tabularx}{\textwidth}{lp{1.2cm}lp{2.7cm}lp{2.7cm}X}
  \toprule
   & Cluster & \multicolumn{2}{l}{CPU/node} & \multicolumn{2}{l}{GPU/node} & \multicolumn{1}{l}{Interconnect} \\
  \midrule
  1 & VSC-5 & 2x & AMD EPYC 7713 \newline 2.0 GHz 33 M cache L3 & 2 x & NVidia A100 \newline 40 GB memory & Infiniband Mellanox HDR \\
  2 & Narval & 2 x & AMD Milan 7413 \newline 2.65 GHz 128M cache L3 & 4 x & NVidia A100 \newline 40 GB memory & Infiniband Mellanox HDR \\
  \bottomrule
  \end{tabularx}
  \caption{Hardware configurations for performance testing.}
  \label{tab:hardware}
\end{table*}

To test our integration of CAF with CUDA Fortran, and to compare it with a hybrid MPI-CUDA Fortran approach, we employed the convolution method for solving potential fields on nested grids in three dimensions \citep{Vorobyov2023ComputingMethod}, referred to from hereon in as CM4NG. This method provides an appropriate benchmark due to its reliance on all forms of shared memory, distributed memory, and GPU parallelism. We performed testing on two different clusters with different hardware configurations. These are the Vienna Scientific Cluster 5 (VSC5)\footnote{\href{https://vsc.ac.at/}{https://vsc.ac.at/}} and Compute Canada's Narval\footnote{\href{https://www.calculquebec.ca/}{https://www.calculquebec.ca/}}\footnote{\href{https://alliancecan.ca/}{https://alliancecan.ca/}}. The specifications of the hardware configurations are shown in Table~\ref{tab:hardware}.

\begin{figure*}[htbp!]
  \centering
  \includegraphics[width=.75\linewidth]{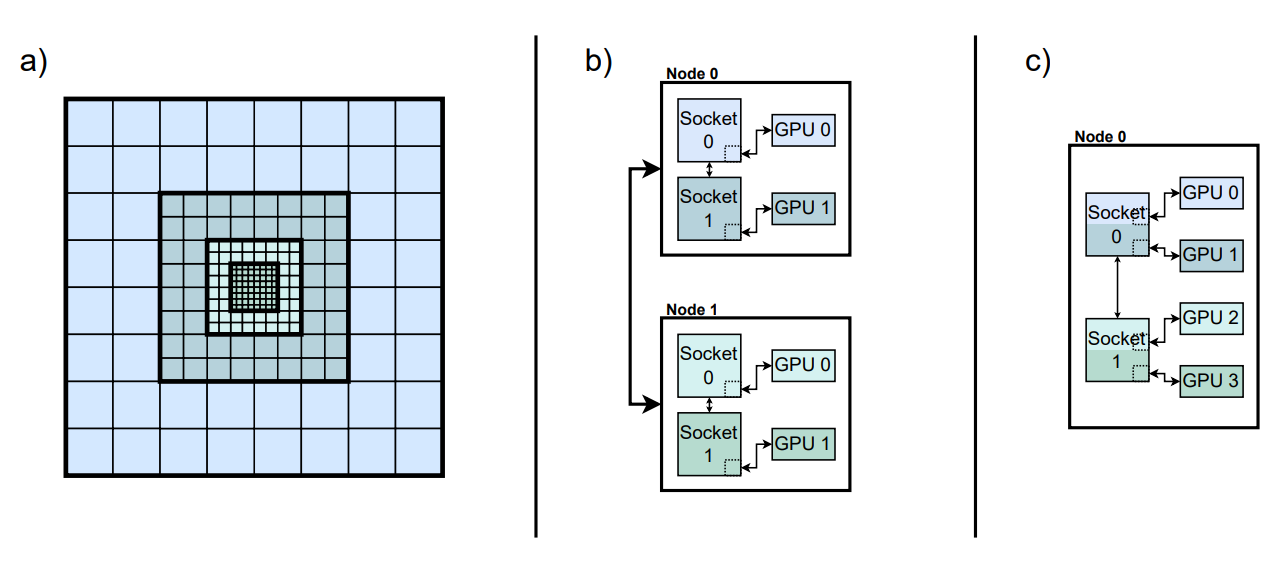}
  \caption{a) Illustration of a two-dimensional slice of the nested grid structure employed in performance testing, showing four nested mesh levels, with each colour corresponding to a new grid level. b) Schematic representation of the grid distribution across four GPUs on VSC5, where each socket has one NUMA node attached to one GPU. c) Schematic representation of the grid distribution across four GPUs on Narval, where each socket has two GPUs attached, each to a different NUMA node. More nested grids require more nodes.}
  \label{fig:grids3}
\end{figure*}

The domain distribution performed on each of the clusters for the convolution theorem on nested grids can be seen in Figure~\ref{fig:grids3}.
In our case, each nested mesh was allocated one GPU, with the number of images per socket varying based on the cluster configuration. GPU utilisation was over 50\% during the test cases, meaning that a computational deceleration would be observed if hosting multiple nested grids on one GPU. If hardware availability is an issue, this may be outweighed by HPC queue times, and a technique called \lq{}CUDA Streams\rq{} could be used to \lq{}split\rq{} the GPUs. However, these were not considered in this study.

In the following text, $N$ is used to refer to the length of the three-dimensional array used by each mesh level in the convolution method. $Ndepth$ is used to denote the number of mesh levels, each one corresponding to a Coarray Fortran (and MPI when used for comparison purposes) image. For these tests, RMA operations in the CH4 device layer using the OFI interface were disabled, as explained above, for both the CAF and MPI tests to ensure comparability. In the following text, CAF+ is used to refer to the fully parallelised hybrid Coarray Fortran, CUDA, and OpenMP approach explained above. MPI+ is similarly used to refer to the aforementioned MPI, CUDA, and OpenMP approach.

\subsection{GPU and distributed memory acceleration}

\begin{figure*}[t]
  \centering
  \includegraphics[width=.98\linewidth]{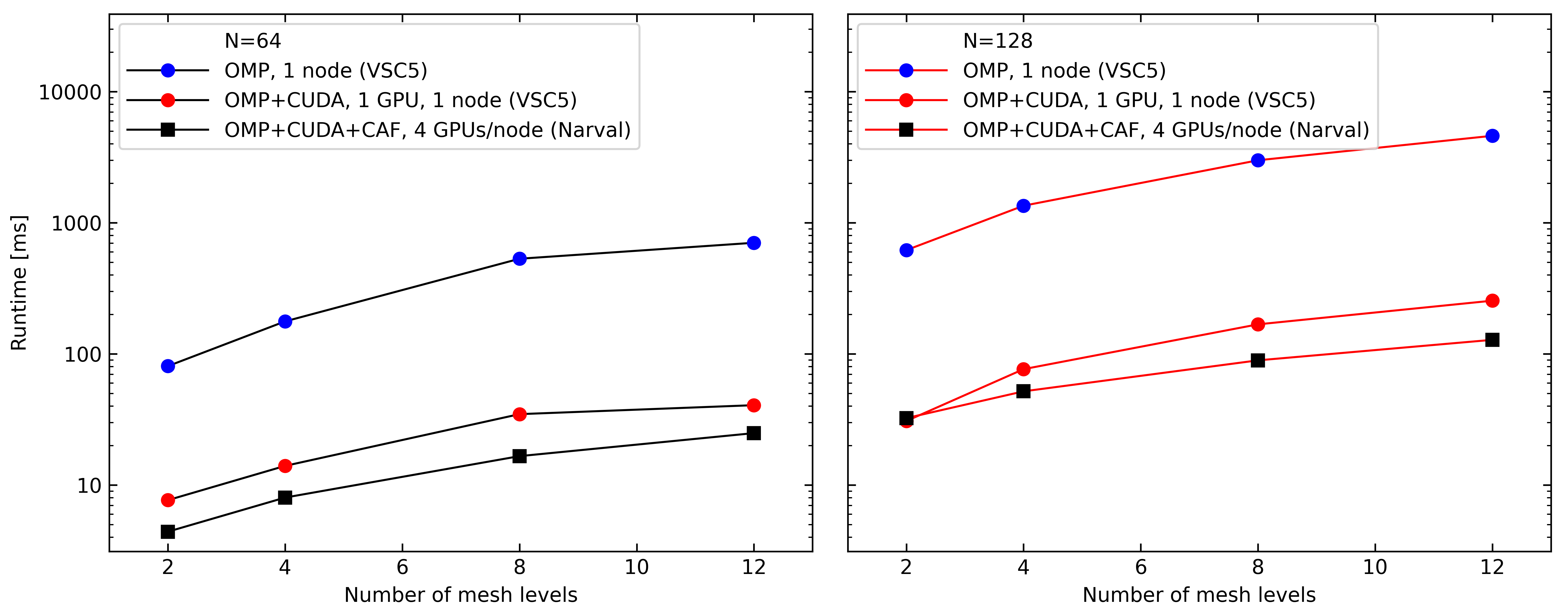}
  \caption{Scaling performance of the convolution method on nested grids when using OpenMP (OMP) parallelisation, OMP combined with CUDA GPU acceleration, and when OMP and CUDA are combined with Coarray Fortran (CAF) distributed memory parallelisation. For this final case of full parallelisation, each nested mesh is assigned one GPU. Left panel: 64$^3$ cells per nested mesh. Right panel: 128$^3$ cells per nested mesh.}\label{fig:res_caf_both}
\end{figure*}

To assess the performance of our integration, we first compared the execution times and scaling efficiency of our solution when only parallelised with OpenMP for shared memory parallelism, with shared memory parallelism combined with a single GPU acceleration, and then compared this with the fully parallelised and optimised Coarray Fortran-CUDA-OpenMP hybrid approach. The results of these tests and how they scale, for both $N=64$ and $N=128$ can be seen in Figure~\ref{fig:res_caf_both} for different numbers of nested levels. It should be noted that this figure uses VSC5 for the single-node results and Narval for the multi-node results. The performance of the two clusters using distributed memory is shown later.


It can be seen that GPU acceleration is highly desirable as it enables acceleration of a factor of approximately 10. Distributed memory parallelism further accelerates our solver by a factor of at least 2.5. In cases such as these, where a potential solver forms part of a larger code which is all Coarray Fortran parallelised, the acceleration is further enhanced by requiring less transfers all to one node to perform the GPU operations. However, such additional complexities are not considered here.

\subsection{CPU-GPU affinity}
\begin{figure}
    \centering
    \includegraphics[width=.8\linewidth]{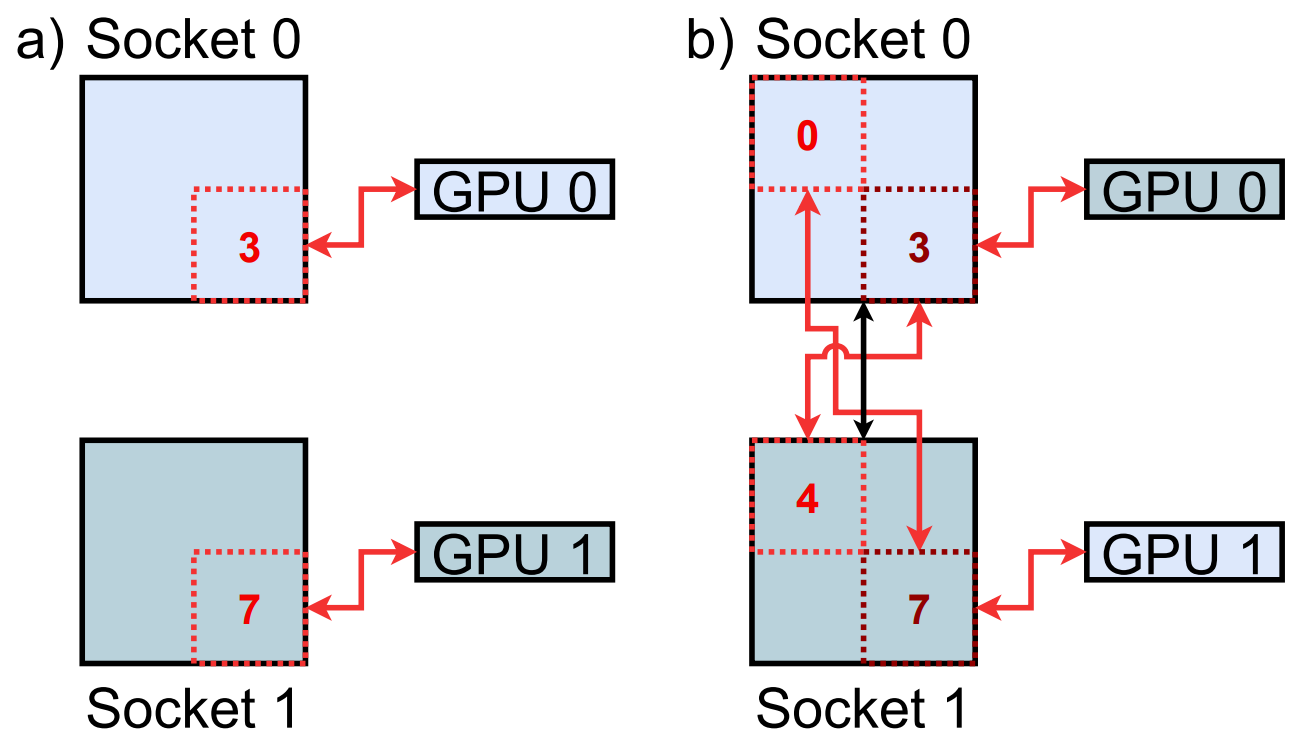}
    \caption{Two configurations for a socket-GPU pair on nodes with two sockets containing NUMA-architecture multiprocessors and two GPUs, where GPUs 0 and 1 are connected to NUMA nodes 3 (socket 0) and 7 (socket 1) respectively. Two processes (light blue and dark blue) are running on the node, one on each socket. a) Shows the optimal configuration of the CPU and GPU processes, where the CPU image runs its GPU tasks on the device directly connected, in this case, process 0 on NUMA node 3 (socket 0) runs on GPU 0, and process 1 on NUMA node 7 (socket 1) runs on GPU 1. b) Shows the worst-case configuration where CPUs perform their GPU calculations on the device connected to the opposite socket, in this case, process 0 on NUMA node 0 (socket 0) runs on GPU 1, and process 1 on NUMA node 4 (socket 1) runs on GPU 0.}
    \label{fig:affinity_graphic}
\end{figure}

To present the importance of CPU-GPU affinity, we confined each nested mesh level to a NUMA node and then performed speed testing by running the code with perfect affinity and perfectly inverse affinity. These optimal and pessimal configurations are illustrated in Figure~\ref{fig:affinity_graphic}. Case a) presents the perfect affinity configuration when the coarray image runs its GPU tasks on the device directly connected to its CPU socket, while case b) shows the worst affinity configuration when the coarray image has to communicate through the other socket to perform its GPU tasks.

\begin{figure}
  \centering
  \includegraphics[width=\columnwidth]{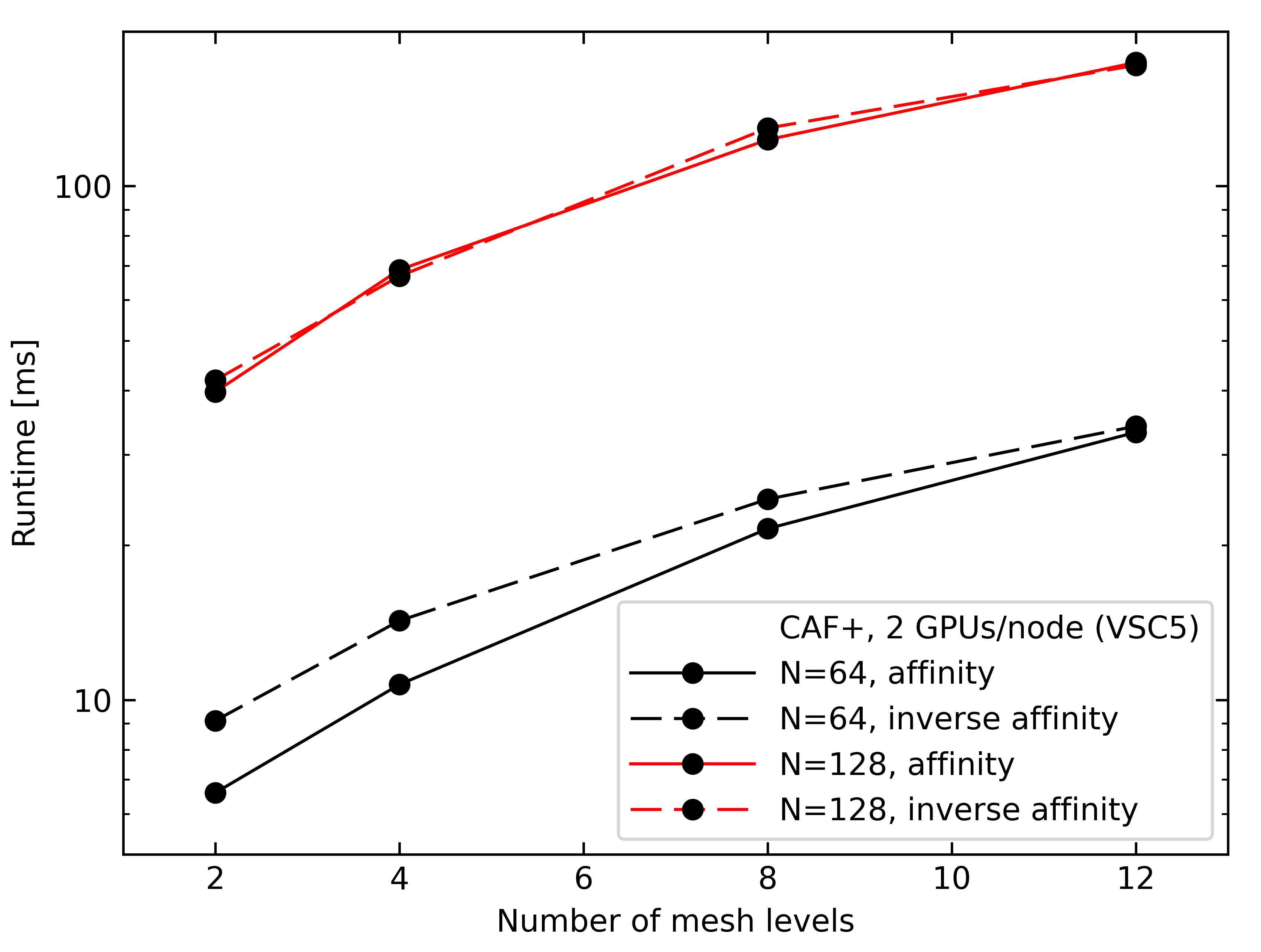}
  \caption{Scaling performance of CM4NG with full parallelisation via OpenMP (OMP), CUDA and Coarray Fortran (CAF), using 2 GPUs per node on VSC5. CPU-GPU affinity is shown for the optimal and pessimal configurations.}\label{fig:res_affinity}
\end{figure}

The results of this testing can be seen in Figure~\ref{fig:res_affinity}. The degree to which CPU-GPU affinity impacts the performance of CM4NG is dependent on the size of the computational task, given that the CPU-GPU transfer time takes up a smaller portion of the solving time as the number of computations increases with grid size. In our case, when it has the most impact, the optimal configuration is approximately 40\% faster than the pessimal configuration. This is seen when using two nested meshes, each with 64$^3$ resolution. For the 64$^3$ resolution cases, the difference between optimal and pessimal configurations saturates, to the point where for 12 nested mesh levels, the advantage is negligible. For the 128$^3$ case, the difference between optimal and pessimal configurations is negligible for all mesh depths.

\subsection{GPUs per node}

\begin{figure}
  \centering
  \includegraphics[width=\columnwidth]{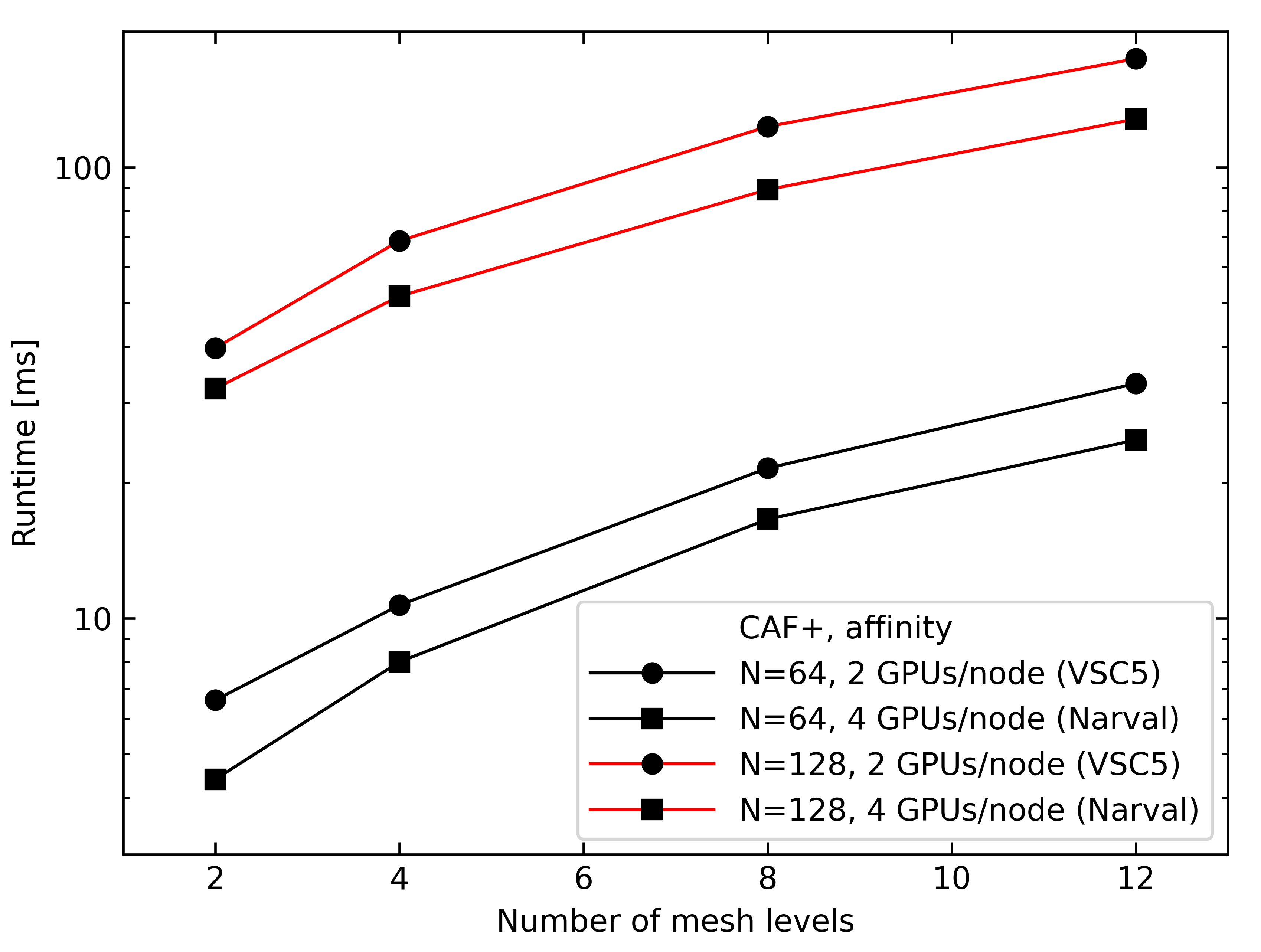}
  \caption{Scaling performance of CM4NG on clusters VSC5 and Narval, using 2 GPUs/node and 4 GPUs/node respectively. CPU-GPU affinity is optimal for these results. CM4NG is configured to allocate each nested mesh level one GPU. The code is fully parallelised with Coarray Fortran, CUDA and OpenMP (CAF+). Both grids with 64$^3$ and 128$^3$ cells per nested mesh level are shown.}\label{fig:res_clusters}
\end{figure}

To test the performance of the code when running on a different number of GPUs per node, we performed scaling tests on VSC5 and Narval. The results of this can be seen in Figure~\ref{fig:res_clusters}. Notable here is the speed-up of the code when using 4 GPUs per node on Narval as opposed to 2 GPUs per node on VSC5. Although Narval's CPUs support a slightly higher clock speed and a larger L3 cache, these differences are mostly due to the fact that inter-node communication is more costly than intra-node communication. When 4 GPUs per node are used as opposed to 2, only 3 nodes in total are needed as opposed to 6. This reduces the transfer time introduced by distributed memory parallelism, as the distribution is across a smaller number of nodes.

\subsection{Coarray Fortran vs MPI}

\begin{figure}
  \centering
  \includegraphics[width=\columnwidth]{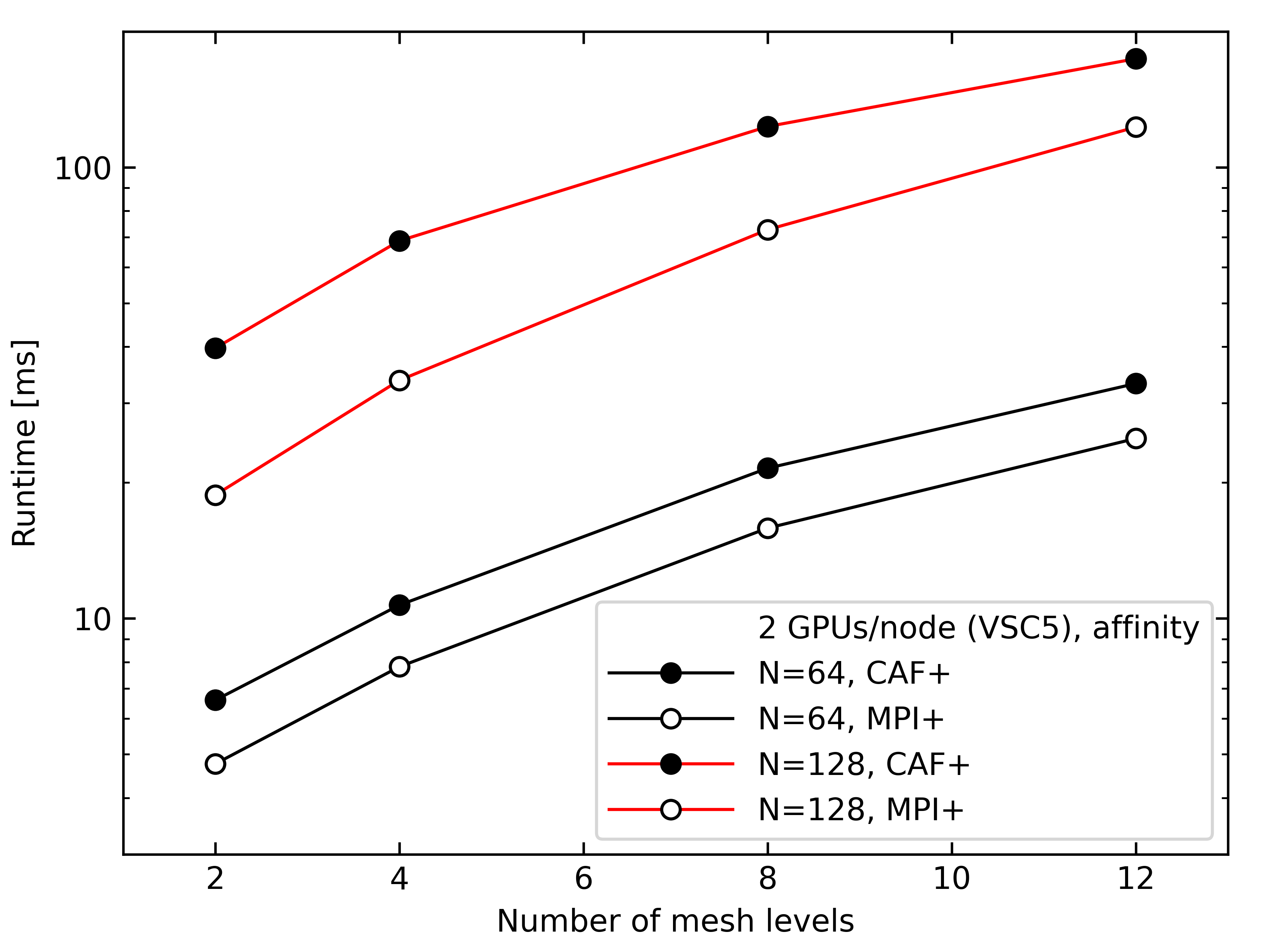}
  \caption{Scaling performance of CM4NG using 2 GPUs per node on VSC5, when CPU-GPU affinity is optimal. Fully parallelised Coarray Fortran, CUDA and OpenMP (CAF+) and MPI, CUDA and OpenMP (MPI+) are compared for both 64$^3$ and 128$^3$ cells per nested mesh level. CM4NG uses one GPU per nested mesh level.}\label{fig:res_vsc5}
\end{figure}

To test the efficiency of the Coarray Fortran implementation explained above, we performed scaling tests for both the CAF and MPI versions of the solver. The results of these tests can be seen in Figure~\ref{fig:res_vsc5}. We observe that the performance of the CAF and MPI versions of the code are comparable, with the MPI version performing approximately 10\% faster for 12 nested mesh levels for the 64$^3$ resolution case, and approximately 5\% faster for the 128$^3$ resolution case. This difference becomes more pronounced as the size of the computational domain decreases, with CAF being 30\% slower for 64$^3$ with 2 nested meshes and 50\% for 128$^3$. This is partly due to the ability of the MPI code to use pinned memory for device transfers, as described above. However, the main reason for this difference is the faster implementation of MPI than CAF by Intel in the \texttt{ifort} compiler. In the cases where the number of mesh levels is lower, the computation forms less than half of the total time to complete the solution, with the majority of time being spent on communication. This means the result is extremely sensitive to the speed of the communication. In real application, for the production of useful results, CM4NG is used at 8 nested mesh levels and above. In cases where data transfers form the majority of the time to complete the solution, and actual computation forms the minority, MPI becomes more desirable.

\subsection{Execution times}

The execution times for the different hardware configurations, as seen, show that Coarray Fortran performs comparably to MPI, scaling in the same manner with no worsening performance as the computational domain increases, both for the nested mesh size and the number of nested meshes. All results demonstrate a near-proportional scaling of the code with nested mesh level. In our method, no asynchronous transfers can be performed so all effects of transfers between the device and the host are reflected in the results. However, as is seen, when more GPUs are clustered on a computational node, the code performs better, given less inter-node transfers are required, which are more costly than intra-node transfers.

We found that for a CM4NG computational domain of a size producing adequate accuracy, the CAF-CUDA-OMP integration showed a 5\% reduction in speed when compared to the MPI-CUDA Fortran approach. This, for us, is considered in the context of the simplicity of implementing Coarray Fortran when compared to MPI. According to our experience, many legacy Fortran codes were written to be run in a serial mode. Such codes can be parallelised with modest efforts to be run on a single node using OpenMP, unless \texttt{common} blocks were utilised, in which case parallelisation is difficult and requires introducing modules. In any case, a distributed memory parallelisation of legacy Fortran codes using MPI requires a significant rewrite of the code which, in terms of the required efforts, may be comparable to writing a new code from scratch. With Coarray Fortran the distributed memory parallelisation of legacy Fortran codes becomes feasible with relatively little effort but significant speed-up. This is because CAF requires introducing only several additional lines and variables to the code, keeping the existing code structure and syntax intact. The fact that Coarray Fortran can be integrated with CUDA Fortran makes the coarray language standard particularly attractive for scientists.

\section{Conclusions}

In this study, we have successfully demonstrated a robust methodology for integrating Intel Coarray Fortran with Nvidia CUDA Fortran, supplemented by OpenMP, to optimise Fortran codes for high-speed distributed memory, GPU, and multiprocessor acceleration. Our approach, focusing on the fusion of these distinct yet powerful paradigms, has shown significant promise in enhancing the computational performance of Fortran codes without the need for extensive rewrites or departure from familiar syntax.

\begin{enumerate}
  \item Performance Improvements: Our results indicate that the Coarray Fortran-CUDA Fortran integration leads to substantial performance gains. We observed, for our use case, only a 5\% reduction in execution time compared to a similar MPI-CUDA Fortran approach. The performance of CAF-CUDA Fortran is significant, considering the comparative ease of implementation of Coarray Fortran over MPI. Our findings underscore the potential of Coarray Fortran, especially for the scientific community that relies heavily on Fortran codes. Its straightforward syntax and implementation make it an accessible and powerful tool for researchers who may not have extensive experience with more complex distributed memory parallelism techniques.
  \item Scalability: The near-linear scaling of our potential solver code with the increase in nested mesh levels and the number of nested meshes highlights the efficiency of our approach, and this scaling is present in both the MPI and Coarray Fortran implementations of the code. This scalability allows the code to be run at a competitive speed on a range of hardware depending on its performance.
  \item Hardware Utilisation: Our methodology's ability to leverage multiple GPUs effectively, as evidenced by improved performance on systems with a higher concentration of GPUs per computational node, points towards the importance of hardware-aware code optimisation and the minimisation of distributed memory transfers. This means the speed of these communications should be as high as possible and not influenced heavily by additional transfers than those to and from the GPU memory. Our methodology avoids this by using pointers, ensuring the most optimal Coarray Fortran and MPI performance.
  \item CPU-GPU affinity: When using Coarray Fortran and CUDA Fortran together, we observe an increase in performance when CPU-GPU affinity is optimised, and show that this is particularly important when data transfer times make up a significant proportion of the complete solution time. We demonstrate the method which can be used to optimise CPU-GPU affinity when using Coarray Fortran via the \texttt{I\_MPI\_PIN\_DOMAIN} environment variable.
  \item Portability: As our approach relies in part on the Intel Fortran compiler, we outline effective solutions to enable the running of Coarray Fortran on AMD processors, the fastest of these being to change the remote memory access protocol used by CAF.
\end{enumerate}

In conclusion, our integration of Coarray Fortran with CUDA Fortran and OpenMP offers a significant step forward in modernizing Fortran-based scientific computing. Multiple implementations are available and are compared and contrasted depending on the use case.

\section*{Acknowledgements}
We are thankful to the referee for the comments and suggestions that helped to improve the manuscript. This work was supported by the FWF project I4311-N27 (J.M., E.I.V.) and RFBR project 19-51-14002 (I.K.).
Benchmarks were performed on the Vienna Scientific Cluster (VSC)\footnote{\href{https://vsc.ac.at/}{https://vsc.ac.at/}} and on the Narval Cluster provided by Calcul Québec\footnote{\href{https://www.calculquebec.ca/}{https://www.calculquebec.ca/}} and the Digital Research Alliance of Canada\footnote{\href{https://alliancecan.ca/}{https://alliancecan.ca/}}.

\bibliographystyle{elsarticle-num} 
\bibliography{references}

\newpage
\appendix
\section{C-Binding in Fortran}
C-binding in Fortran is a powerful feature that allows interoperability with C, enabling the use of C data types, and the calling of C functions and libraries. This feature is particularly useful in high-performance computing where leveraging both Fortran's computational efficiency and C's extensive library ecosystem can be advantageous. 

In the context of our study, C-binding plays a critical role in enabling robust interaction between the \texttt{nvfortran} and \texttt{ifort} compilers, allowing them to communicate according to the C standard. When implementing pure Fortran interfaces between the compilers, we met numerous segmentation errors which were avoided when communication was facilitated by binding the connection with C. This section provides an overview of how C-binding is used in our methodology and a simple example for illustration.

\subsection{Using C-binding in Fortran to combine Fortran compilers and allow a robust common memory space}
In the case below, code compiled with compiler $A$ will receive the location of a variable in memory compiled by compiler $B$. The code compiled by compiler $A$ will then be able to access the variable directly.

To use C-binding in Fortran, specific steps must be followed:

\begin{enumerate}
  \item[B1] \textbf{Define the subroutine to return the memory location}: In code compiled with compiler $B$, define a subroutine with the \texttt{bind(C)} attribute, which returns the C address of a variable.

  \item[A1] \textbf{Define the interface to the subroutine}: In the code compiled with compiler $A$, define an interface to the previous subroutine, also using the \texttt{bind(C)} attribute. This must be done inside a module.

  \item[A2] \textbf{Call the subroutine}: In the code compiled with compiler $A$, call the subroutine defined in step B1. This will return the C address of the variable to the code compiled with compiler $A$.

  \item[A3] \textbf{Convert the C address to a Fortran pointer}: In the code compiled with compiler $A$, convert the C address to a Fortran pointer. This can be done using the \texttt{c\_f\_pointer} intrinsic function.
\end{enumerate}

\subsection{Example of C-Binding for memory location passing}
Below is a simple example demonstrating the use of C-binding in Fortran to pass the memory location of a variable from one compiler to another.

\subsubsection{Code compiled with compiler B}
\begin{verbatim}
! module containing the variable
module my_vars
  implicit none
  integer, target :: a
end module my_vars
  
! subroutine returning the C address
subroutine get_a(value) bind(C, name="get_a")
  use iso_c_binding, only: c_ptr, c_loc
  use my_vars, only: a
  implicit none
  type(c_ptr), intent(out) :: value
  value = c_loc(a)
end subroutine get_a
\end{verbatim}

\subsubsection{Code compiled with compiler A}
\begin{verbatim}
! interface to subroutine
module interface_compilers
  use iso_c_binding, only: c_ptr
  implicit none

  interface
    subroutine get_a(value) bind(C, name="get_a")
      type(c_ptr), intent(out) :: value
    end subroutine get_a
  end interface
end module interface_compilers

! call the subroutine and convert to pointer
program main
  use iso_c_binding, only: c_ptr
  use interface_compilers
  implicit none

  type(c_ptr) :: a_c_ptr
  integer, pointer :: a_ptr

  call get_a(a_c_ptr)
  call c_f_pointer(a_c_ptr, a_ptr)

  ! use as normal Fortran variable
end program main
\end{verbatim}

In this example, the Fortran program defines a module \texttt{interface\_compilers} containing an interface to a C function, written in Fortran, named \texttt{get\_a}. This methodology can be used to not only pass C addresses but more generally call functions and subroutines across Fortran code compiled with different compilers.

\subsection{Integration in Our Methodology}
In our methodology, C-binding is employed to create interfaces between CUDA Fortran and Coarray Fortran components. This is crucial for transferring data and control between different parts of the application, which are compiled with different compilers. By using C-binding, we ensure a robust and efficient interaction between these components.

\end{document}